\documentstyle[12pt]{article}
\begin{document}
\begin{center}
{\Large \bf{The lowest-order short-distance contribution to the 
 $B_{s}\rightarrow\gamma\gamma$ decay}}
\  \par
\  \par
\   \par
\    \par
\   \par
{\Large \bf {Gela G.~Devidze}}\footnote{
Permanent address: High Energy Physics Institute, Tbilisi State 
University, University St.9, 
Tbilisi 380086, Rep. of Georgia. E-mail:devidze@hepi.edu.ge}\\
\    \par
\    \par
\    \par
\    \par
Department of Physics, University of Durham, Durham DH1 3LE, UK\\
devidze@hepi.edu.ge~~~~~~~~~devidze@hep.phys.soton.ac.uk\\
\end{center}
\ \par
\ \par
\   \par
\   \par
{\bf Abstract:} The complete calculation of the lowest order short-distance 
contributions to the $B_{s}\rightarrow\gamma\gamma$ decay in the SM are 
presented. The amplitude and branching ratio are calculated.
\  \par
\   \par
\   \par
\newpage

The theoretical and experimental investigations of rare $B$-meson's decays 
provide precise test of the Standard Model (SM) and possible new physics beyond. 
Among the rare $B$ decasys with particularly clean experimental signature is 
$B_{s}$-meson two photon radiative decay $B_{s}\rightarrow\gamma\gamma$. The 
present experimental bound on this decay is [1]
$$
	Br(B_{s}\rightarrow\gamma\gamma)~<~1.48\cdot10^{-4}  \eqno{(1)}
$$

	$B$-meson double radiative decay has rich final state. Two photon can be in a 
$CP$-odd and $CP$-even state. Tferefore this decay allows us to study $CP$ 
violating effects. In the SM the branching ratio of $B_{s}\rightarrow\gamma\gamma$ 
decay is of order$10^{-7}$ without QCD corrections [2-5]. The branching ratio of 
this decay is enhanced with the addition of the QCD corrections [6-14]. The QCD 
corrections may correct the lowest order short-distance contributions to the
$B_{s}\rightarrow\gamma\gamma$ decay in order of magnitude\footnote{In the paper 
[11] the authors have estimated the long-distance contributions to the 
$B_{s}\rightarrow\gamma\gamma$ decay arising from charmed-meson intermediate
states. They have obtained that contributions of the diagrams with $D^{*}_{s}$ 
may enhance the branching ratio more than an order of magnitude. The authors have 
mentioned that they neglected quite a few possible contributions to the process. 
They hope that the detail investigation does not invalidate the results presented 
in the paper [11].}.

	The planed experiments at the upcoming SLAC and KEK $B$-factories and hadronic
accelerators are capable to measure the branching ratio as low as $10^{-8}$. Therefore 
one expects the double radiative decay of the $B_{s}$-meson 
$B_{s}\rightarrow\gamma\gamma$ 
to be seen in these future facilities, thus stimulate theoretical investigations.

	This decay is sensitive to possible new physics beyond the SM. Interstingly, the 
branching ratio can be enhanced in extensions of the SM [15,16]. Before one goes on to 
study other new physics which potentially can influence this decay, it stands to reasons
to improve upon previous calculations [2-5].

	In this paper we study the lowest-order short-distance contributions to the 
$B_{s}\rightarrow\gamma\gamma$ decay in the SM without QCD corrections. We do not
neglect mass of $s$-quark. It is not immediately obvious how such investigation correct 
the branching ratio. The diagrams contributing to this decay are presented in Fig.1. The 
lowest-order short-distance contribution to the $B_{s}\rightarrow\gamma\gamma$ decay 
arise from the following set of graphs: i) triangle diagrams with external photon leg (one 
particle reducible (OPR) diagrams), ii) box diagrams (one particle irreducible (OPI) 
diagrams).

	One can write doun the amplitude for the decay $B_{s}\rightarrow\gamma\gamma$ in
the following form, which is correct after gauge fixing for final photons
$$
   T(B_{s}\rightarrow\gamma\gamma) = 
         \epsilon^{\mu}_{1}(k_{1})\epsilon^{\nu}_{2}(k_{2}) 
 [Ag_{\mu\nu} + iB\epsilon_{\mu\nu\alpha\beta}k^{\alpha}_{1}k^{\beta}_{2}]  \eqno{(2)}
$$
where $\epsilon_{1}^{\mu}(k_{1})$ and $\epsilon_{2}^{\nu}(k_{2})$ are
the polarization vectors of final photons with momenta $k_{1}$ and $k_{2}$
respectively. 
	Let us fix photon polarization by the conditions
$$
	\epsilon_{i}\cdot\epsilon_{j} = 0,~~~~~~~~~i,j = 1,2   \eqno{(3)}
$$

	The conditions (3) with allowance for the energy-momentum conservation in the 
diagrams of Fig.1 yeld
$$
	\epsilon\cdot P = \epsilon\cdot p_{b} = \epsilon\cdot p_{s} = 0  \eqno{(4)}
$$
where
$$
	P = k_{1} + k_{2},~~~~~~~~p_{b} = p_{s} + k_{1} + k_{2}   \eqno{(5)}
$$
Formulae (3)-(5) lead to useful kinematikal relation
$$
	k_{1}\cdot k_{2} = P\cdot k_{i} = \frac12 M^{2}_{B_{s}},~~~~~
	P\cdot p_{b} = m_{b}M_{B_{s}},~~~~P\cdot p_{s} = - m_{s}M_{B_{s}}
$$
$$
	p_{b}\cdot p_{s} = - m_{s}m_{b},~~~~~~~~p_{b}\cdot k_{i} = 
  \frac12 m_{b}M_{B_{s}},~~~~~p_{s}\cdot k_{i} = - \frac12 m_{s}M_{B_{s}}
     \eqno{(6)}
$$

	With the aid of (3)-(6) one can calculate the cobntribution of 
each diagrams to the amplitude $T$. We used the 't Hooft-Feynman gauge
 and evaluated divergent Feynman integrals by means of dimensional 
regularization. Only OPR diagrams contain divergent parts. The 
divergent parts mutually cancel in the sum of amplitude and due to the 
GIM mechanism [17].

Using formula (2) we directly obtain the expression for the branching ratio
$$
Br(B_{s} \rightarrow \gamma\gamma)=\frac{1}{32 \pi M_{B_{s}}\Gamma_{tot}}
[4\mid A \mid^{2} + \frac{1}{2}M_{B}^{4}\mid B \mid^{2}] \eqno{(7)}
$$
As from Fig.1 is seen the correct procedure assumes the necessity of final photon 
rearangament. In the kinemastics (3)-96) this procedure leads to doubling of 
all contributions exept of diagrams 19 and 20, where both photons are emitted
from the same space-time point:
$$
A=A_{19} +A_{20} +2{\sum_{i=1}^{34}}^{'}A_{i}, \hspace{1cm}
B=B_{19} +B_{20} +2{\sum_{i=1}^{34}}^{'}B_{i},   \eqno{(8)}
$$
   where the stress over the sum means absence in the sum of 19-th
and 20-th terms.

    The amplitude $T(B_{s} \rightarrow \gamma\gamma)$ and hence its
$CP$-even and $CP$-odd parts can be written as a sum of contributions
from up-quarks
$$
T(B_{s}\rightarrow\gamma\gamma) =
\sum_{i=u,c,t}\lambda_{i}T_{i}=\lambda_{u}T_{u}+\lambda_{c}T_{c}+
\lambda_{t}T_{t},     \eqno{(9)}
$$
where $\lambda_{i} = V_{is}V^{*}_{ib}$ ($V_{kl}$ being the corresponding 
elements of CKM matrix). Using the unitarity of the CKM matrix 
($\sum\lambda_{i} = 0$) one can rewrite it in the form
$$
T=\lambda_{t}\{T_{t}-T_{c}+\frac{\lambda_{u}}{\lambda_{t}}(T_{u}-T_{c})\}
         \eqno{(10)}
$$
Below we restrict ourselves to evaluating the amplitude in the leading order
($1/M^{2}_{W}$). The $u$-quark and $c$-quark contributions are equal in this 
approximation ($T_{u} = T_{c}$). So, the expression for the amplitude becomes 
a simpler form
$$
	T = \lambda_{t}(T_{t} - T_{c})    \eqno{(11)}
$$

	Only the OPR diagrams have nonzero contributions into amplitude $A$ in this
approximation. As conserning the amplitude $B$, it is grathered both from OPR 
diagrams and OPI diagrams 34 of Fig.1. The corresponding contributions are
$$
	A = i\frac{\sqrt2}{32\pi^{2}} G_{F}f_{B}(m_{b} - m_{s})M_{B_{s}}\lambda_{t}
\{ (\frac{m_{b}}{m_{s}} + \frac{m_{s}}{m_{b}})[C(x_{t}) - C(x_{c})] + 
 C_{1}(x_{t}) - C_{1}(x_{c})\}
$$
$$
B = i\frac{\sqrt2}{16\pi^{2}} G_{F}f_{B}\lambda_{t}
\{ (\frac{m_{b}}{m_{s}} + \frac{m_{s}}{m_{b}})[C(x_{t}) - C(x_{c})] + 
 C_{2}(x_{t}) - C_{2}(x_{c}) - 32M^{2}_{B_{s}}I(m^{2}_{c}) \}   \eqno{(12)}
$$
where
$$
	C(x) = \frac{22x^{3} - 153x^{2} + 159x - 46}{6(1 - x)^{3}} + 
              \frac{3(2 - 3x)x^{2}\ell nx }{(1 - x)^{4} }
$$

$$
	C_{1}(x) = \frac43 \cdot \frac{6x^{3} - 27x^{2} + 25x - 9 + 6x^{2}\ell nx}
                                     {(1 - x)^{3}} 
$$
$$
	C_{2}(x) = \frac{22x^{3} - 12x^{2} - 45x + 17}{3(1 - x)^{3}} + 
              \frac{2x(8x^{2} - 15x + 4)\ell nx }{(1 - x)^{4} }
$$
$$
   I(m_{c}^{2}) = - \frac{1}{2M^{2}_{B_{s}}} \{ 1 + \frac{m^{2}_{c}}
{M^{2}_{B_{s}}}(\ell n^{2}\frac{1 + \beta }{1 - \beta } - \pi^{2} - 
 2i\pi\ell n\frac{1 + \beta }{1 - \beta}) \}
$$
$$
   x_{t} = \frac{m^{2}_{t}}{M^{2}_{W}},~~~~~~
 \beta = \sqrt{1 - 4\frac{m^{2}_{c}}{M^{2}_{B_{s}}}}     \eqno{(13)}
$$
We also used the following relations for hadronic matrix elements

$$
<0\mid\bar s\gamma_{\mu}\gamma_{5}b\mid B_{s}(P)>=-if_{B}P_{\mu},~~~~~
<0\mid\bar s\gamma_{5}b\mid B_{s}(P)> \approx if_{B}M_{B_{s}} \eqno{(14)}
$$
Using expressions (7),(12) and (13) one can estimate the branching ratio of the 
$B_{s}\rightarrow\gamma\gamma$ decay
$$
	Br(B_{s}\rightarrow\gamma\gamma) = 2\cdot 10^{-7}    \eqno{(15)}
$$
We have used the following set of parameters: $m_{t}$ = 175~GeV, $m_{b}$ = 4.8~GeV, 
$m_{s}$ = 0.5~GeV, $f_{B}$ = 200~MeV, $\lambda_{t} = 4\cdot 10^{-2}$, 
$M_{B_{s}}$ = 5.3~GeV, $\Gamma_{tot}(B_{s}) = 5\cdot 10^{-4}$~eV. It should be 
mentioned that we do not neglect mass of $s$-quark. If one neglect mass of $s$-quark 
the branching ratio becomes $30\%$ larger than the result (15). The upcoming $B$ 
factories at SLAC, KEK and hadronic $B$ projects at LHC, HERA, TEVATRON will be 
possible to study decay modes with branching ratio as small as $10^{-8}$. Branching 
ratio $10^{-7}$ will be mesurable in these facilities. Detail investigation of the 
lowest-order short-distance contributions to the $B_{s}\rightarrow\gamma\gamma$ decay
deckreases the branching ratio. This decay is sensitive to parameters and requierst 
further experimental and theoretical investigations.
\  \par
\  \par
\  \par
{\Large \bf {Acknowledgments}}\\
This research was supported in part by The Royal Society. 
I am very grateful to prof. A.D.~Martin for warm hospitality. I also would
like to thank G.R.~Jibuti, A.G.~Liparteliani for discussions.

\newpage
\begin{center}
Figure
\end{center}
\  \par
\  \par
\  \par
\  \par
\  \par
Fig.1. One particle reducible and one particle irreducible diagrams 
contributing to the $B_{s}\rightarrow\gamma\gamma$ decay.

\end{document}